\documentclass[aps,prl,twocolumn,preprintnumbers,groupedaddress]{revtex4}

\usepackage{todonotes}

\usepackage{amssymb}
\usepackage{epsfig} 
\usepackage{amsmath,color} 
\usepackage{graphicx} 
\usepackage{ulem} 
\usepackage{multirow}
\usepackage{tabularx}
\usepackage{slashed}

\usepackage{wrapfig}

\usepackage{tikz}
\usetikzlibrary{positioning,arrows}
\usetikzlibrary{decorations.pathmorphing}
\usetikzlibrary{decorations.markings}

\newcommand{\beq}{\begin{equation}}
\newcommand{\eeq}{\end{equation}}
\newcommand{\bea}{\begin{eqnarray}}
\newcommand{\eea}{\end{eqnarray}}
\newcommand{\ba}{\begin{array}}
\newcommand{\ea}{\end{array}}
\newcommand{\bi}{\begin{itemize}}
\newcommand{\ei}{\end{itemize}}
\newcommand{\bn}{\begin{enumerate}}
\newcommand{\en}{\end{enumerate}}
\newcommand{\bc}{\begin{center}}
\newcommand{\ec}{\end{center}}
\renewcommand{\l}{\left}
\renewcommand{\r}{\right}

\newcommand{\eqs}[2]{Eqs.~(\ref{#1}) and (\ref{#2})}

\newcommand{\eV}{\mathinner{\mathrm{eV}}}

\newcommand{\MeV}{\mathinner{\mathrm{MeV}}}

\begin{document}

\preprint{FTUV-16-09-27}
\preprint{IFIC/16-67}

\title{Impact of CP-violation on neutrino lepton number asymmetries revisited}


\author{Gabriela Barenboim$^1$}
\email[]{Gabriela.Barenboim@uv.es}


\author{Wan-Il Park$^1$}
\email[]{Wanil.Park@uv.es}
\affiliation{$^1$ Departament de F\'isica Te\`orica and IFIC, Universitat de Val\`encia-CSIC, E-46100, Burjassot, Spain}


\date{\today}

\begin{abstract}
We revisit the effect of the (Dirac) CP-violating phase on neutrino lepton number asymmetries in both  mass- and flavor-basis.
We found that, even if there are sizable effects on muon- and tau-neutrino asymmetries, the effect on the asymmetry of electron-neutrinos is at most similar to the upper bound set by BBN for initial neutrino degeneracy parameters smaller than order unity.
We also found that, for the asymmetries in mass-basis, the changes caused by CP-violation is of sub-\% level which is unlikely to be accesible neither in the current nor in the forthcoming experiments.
\end{abstract}

\pacs{}

\maketitle


\section{Introduction}

If the three active neutrinos are Dirac particles, all the possible sources of CP violation are restricted to only one so-called Dirac phase in the Pontecorvo-Maki-Nakagawa-Sakata (PMNS) matrix \cite{Pontecorvo:1957cp,Maki:1962mu}.
This phase can cause an asymmetry between conversion probabilities associated with a pair of neutrino flavors and their anti-particles \cite{Bilenky:1980cx,Cabibbo:1977nk,Barger:1980jm}. Being this phase also present in the case neutrino are Majorana particles, it is not possible  to distinguish the nature of neutrinos through oscillation experiments. 
Experimental sensitivities are not good enough yet to determine the phase, and all values of the CP-phase ranging $0-2\pi$ are still allowed. Future longbaseline experiments will measured it eventually (see for example Refs.~\cite{Bernabeu:2010rz}).

If lepton number asymmetries of neutrinos exist, their CP-violating phase can change the prediction of the late time lepton number asymmetries established after the flavor mixing reached an equilibrium before Big Bang Nucleosynthesis (BBN) \cite{Gava:2010kz}. 
If the initial neutrino lepton number asymmetries in the very early universe arelarge enough, CP-violation may cause a sizable change of the asymmetry of electron-neutrinos which is constrained by BBN, and a change of $\Delta N_{\rm eff}$ the number of extra neutrino species. 
As a result, it might be possible for a non-zero CP-phase to constrain the initial configuration of neutrino asymmetries (or the total lepton number asymmetry).

Recently, it has been shown that, even if Big Bang Nucleosynthesis (BBN) constrains the lepton number asymmetry of electron-neutrinos quite tightly such as $|L_e| \lesssim 10^{-3}$, muon- and tau-neutrinos can still have large asymmetries which can result in $\Delta N_{\rm eff} \sim \mathcal{O}(0.1-1)$ \cite{Barenboim:2016shh}.
It has been also pointed out that a correct estimation of $\Delta N_{\rm eff}$ coming from asymmetric neutrinos should be done using neutrino mass-eigenstates instead of flavor-eigenstates \cite{Barenboim:2016lxv}.
However, in those works, the Dirac CP-violating phase was set to zero for simplicity without discussing its possible impact on the eventual asymmetries of neutrinos. 
The effect of CP-violation on asymmetries of neutrino flavors depends on the configuration and sizes of the initial asymmetries, and it is not clear how $\Delta N_{\rm eff}$ measured in mass-basis would be affected.
Hence, for large asymmetries leading to $\mathcal{O}(0.1-1)$ extra neutrino species that might be helpful for a better fit of cosmological data, it is worth to revisit the effect of CP-violation on the neutrino asymmetries.

In this paper, we revisit the effect of CP-violation on neutrino lepton number asymmetries in both flavor- and mass-basis by solving numerically the quantum kinetic equations (QKEs) of neutrino/anti-neutrino density matrices.

\section{Quantum kinetic equations with CP-violation}

Lepton number asymmetries of neutrinos can be defined as 
\beq
\mathbf{L}_{\rm f} = \frac{\rho-\bar{\rho}}{n_\gamma}
\eeq
where $\rho/\bar{\rho}$ and $n_\gamma$ are the density matrix of neutrino/anti-neutrino and the number density of photons, respectively.
For a mode of momentum $p$, the Fourier component of $\rho/\bar{\rho}$ can be expressed in terms of polarization vectors $\mathbf{P}/\bar{\mathbf{P}}$ and Gell-Mann matrices $\lambda_i \ (i=1-8)$ as
\beq
\rho_p = \frac{1}{3} \sum_{i=0}^8 P_i \lambda_i, \quad \bar{\rho}_p = \frac{1}{3} \sum_{i=0}^8 \bar{P}_i \lambda_i
\eeq   
where $\lambda_0$ is the $3\times3$ identity matrix.
The evolution equations of $\rho_p$ and $\bar{\rho}_p$ are given by \cite{Sigl:1992fn,Pantaleone:1992eq}
\bea 
\label{eom-rho}
i \frac{d\rho_p}{dt} &=& \l[ \Omega + \sqrt{2} G_F \l( \rho - \bar{\rho} \r), \rho_p \r] + C \l[ \rho_p \r] \\
\label{eom-rhobar}
i \frac{d \bar{\rho}_p}{dt} &=& \l[ -\Omega + \sqrt{2} G_F \l( \rho - \bar{\rho} \r), \rho_p \r] + C \l[ \bar{\rho}_p \r]
\eea 
In the above equations,
\beq \label{Omega}
\Omega = \frac{\mathbf{M}^2_{\rm f}}{2 p} - \frac{8 \sqrt{2} G_F p \mathbf{E}_\ell}{3 m_W^2}
\eeq
where $\mathbf{M}^2_{\rm f}$ is the mass-square matrix of neutrinos in flavor-basis, $G_F$ the Fermi constant, $m_W$ the mass of $W$-boson, $\mathbf{E}_\ell = {\rm diag}(E_{ee}+E_{\mu\mu}, E_{\mu\mu},0)$ the energy density of charged leptons, $\rho = (1/2 \pi^2) \int_0^\infty \rho_p p^2 dp$ (and similarly for $\bar{\rho}$), and $C[\dots]$ is the collision term.
We take $C[\rho_p] = - i D_{\alpha \beta} [\rho_p]_{\alpha \beta}$ for $\alpha \neq \beta$ only, and similarly for $C[\bar{\rho}_p]$ \cite{Dolgov:2002ab}.
Also, the initial condition for the evolution equations \eqs{eom-rho}{eom-rhobar} can be set as
\beq \label{rho-p-initial}
\rho_p = f(y,0)^{-1} {\rm diag}(f(y,\xi_e), f(y,\xi_\mu), f(y,\xi_\tau)), 
\eeq
and similarly for $\bar{\rho}_p$ but with $\xi_\alpha \to -\xi_\alpha$, where $f(y,\xi_\alpha)=\l( e^{y-\xi_\alpha}+1 \r)^{-1}$ is the occupation number for a mode $y \equiv p/T$ with $\xi_\alpha$ being the initial degeneracy parameter of $\nu_\alpha$.

The evolution of the density matrices is governed by various effects as shown in \eqs{eom-rho}{eom-rhobar}.
Hence, generically, it is difficult or highly non-trivial to predict analytically the asymmetries at the final equilibrium.
However, it may be possible to get a hint by resorting to two properties of the evolution:
(i) The evolution of a mode near the average momentum in the absence of the neutrino self-interaction term (i.e., the terms of ``$G_F (\rho-\bar{\rho})$'' in \eqs{eom-rho}{eom-rhobar}) can mimic the collective behavior of $\rho$ (or $\bar{\rho}$) including the self-interaction term \cite{Abazajian:2002qx},
(ii) Since all the contributions other than vacuum one in the right-hand side of \eqs{eom-rho}{eom-rhobar} eventually diminish and become negligible, at the final equilibrium the shape of density matrices should be determined by the vacuum contribution.
These properties imply that the structure of $\mathbf{M}^2_{\rm f}$ may provide an idea of the eventual configuration of lepton number asymmetries.
For a mass-square matrix of neutrinos ($\mathbf{M}^2_{\rm m}  = {\rm diag}(m_1^2, m_2^2, m_3^3)$) in mass-basis, the corresponding one in the flavor basis is given by, 
\beq
\mathbf{M}_{\rm f}^2 = U_{\rm PMNS} \mathbf{M}_{\rm m}^2 U_{\rm PMNS}^\dag
\eeq
where the PMNS matrix is 
\begin{widetext}
\beq
U_{\rm PMNS} = 
\l( \begin{array}{ccc}
1 & 0 & 0 \\
0 & c_{23} & s_{23} \\
0 & - s_{23} & c_{23} 
\end{array} \r)
\l( \begin{array}{ccc}
c_{13} & 0 & s_{13} e^{-i \delta} \\
0 & 0 & 0 \\
- s_{13} e^{i \delta} & 0 & c_{13} 
\end{array} \r)
\l( \begin{array}{ccc}
c_{12} & s_{12} & 0 \\
- s_{12} & c_{12} & 0 \\
0 & 0 & 1
\end{array} \r)
\eeq
\end{widetext}
with $s_{ij}/c_{ij}=\sin\theta_{ij}/\cos\theta_{ij}$ and $\delta$ being the Dirac CP-violating phase.
The entries of $\mathbf{M}_{\rm f}^2$, denoted as $m^2_{\alpha \beta}$, are found to be
\begin{widetext}
\bea
\label{m2ee}
m_{ee}^2 &=& m_3^2 - c_{13}^2 \Delta m_{32}^2 - c_{12}^2 c_{13}^2 \Delta m_{21}^2 
\\
m_{\mu \mu}^2 &=& m_3^2 - \l( c_{23}^2+s_{13}^2 s_{23}^2 \r) \Delta m_{32}^2 - \l( s_{12}^2 c_{23}^2 + c_{12}^2 s_{13}^2 s_{23}^2 \r) \Delta m_{21}^2 - 2 c_{12} s_{12} s_{13} c_{23} s_{23} c_\delta \Delta m^2_{21}
\\
m_{\tau\tau}^2 &=& m_3^2 - \l( s_{23}^2+s_{13}^2 c_{23}^2 \r) \Delta m_{32}^2 - \l( s_{12}^2 s_{23}^2 + c_{12}^2 s_{13}^2 c_{23}^2 \r) \Delta m_{21}^2 + 2 c_{12} s_{12} s_{13} c_{23} s_{23} c_\delta \Delta m^2_{21} 
\\
m_{e\mu}^2 &=& c_{12} s_{12} c_{13} c_{23} \Delta m^2_{21} + c_{13} s_{13} s_{23} \l( \Delta m^2_{32} + c_{12}^2 \Delta m_{21}^2 \r) e^{- i\delta}
\\
m_{e\tau}^2 &=& - c_{12} s_{12} c_{13} s_{23} \Delta m^2_{21} + c_{13} s_{13} c_{23} \l( \Delta m^2_{32} + c_{12}^2 \Delta m_{21}^2 \r) e^{- i\delta}
\\
\label{m2mutau}
m_{\mu\tau}^2 &=& c_{23} s_{23} \l[ c_{13}^2 \Delta m^2_{32} + \l( s_{12}^2 - c_{12}^2 s_{13}^2 \r) \Delta m^2_{21} \r] + c_{12} s_{12} s_{13} \Delta m^2_{21} \l( s_{23}^2 e^{i \delta} - c_{23}^2 e^{-i \delta} \r)
\eea
\end{widetext}
where $\Delta m^2_{ij} \equiv m^2_i-m^2_j$.
From Eqs.~(\ref{m2ee})-(\ref{m2mutau}), one can see that for the measured central values of $\theta_{ij}$ and $\Delta m^2_{ij}$ the effect of a non-zero CP-phase is driven dominantly by the $\delta$-dependence of $m^2_{e\mu}$ and $m^2_{e\tau}$.
It is worth to note that,
if $\delta = (\pi/2, 3\pi/2)$ and $\theta_{23} = \pi/4$,
\beq \label{msq-eq}
m^2_{\mu\mu} = m^2_{\tau\tau}
\eeq
and
\beq \label{CPV-equal}
{\rm Im}[m^2_{e\mu}] = {\rm Im}[m^2_{e\tau}] \gg {\rm Re}[m^2_{e\mu}] = - {\rm Re}[m^2_{e\tau}]
\eeq
As discussed in Ref.~\cite{Barenboim:2016shh}, for $\delta = (0, \pi)$ if $\theta_{23}=\pi/4$ and either $\theta_{12}$ or $\theta_{13}$ are zero,
$m^2_{\mu\mu}=m^2_{\tau\tau}$ and $|{\rm Re}[m^2_{e\mu}]|=|{\rm Re}[m^2_{e\tau}]|$ which results in $L_\mu=L_\tau$.
Similarly, \eqs{msq-eq}{CPV-equal} may be hinting towards $L_\mu=L_\tau$ again or at least a tendency to such an equalization as $|\delta| \to \pi/2$. 
On the other hand, it is difficult to get an insight on the dependence of $L_e$ on $\delta$ from the structure of $\mathbf{M}^2_{\rm f}$.

\section{Numerical results}

\begin{figure}[ht!]
\begin{center}
\includegraphics[width=0.47\textwidth]{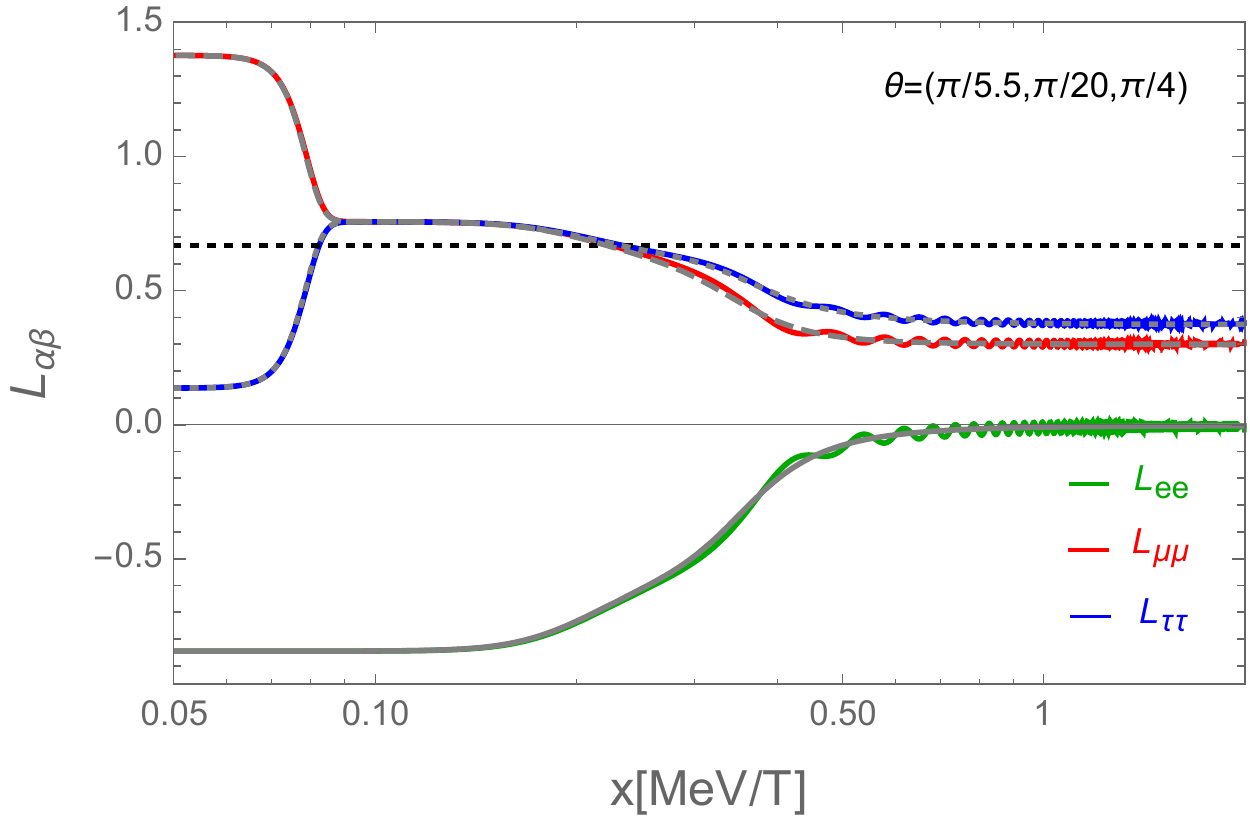}
\caption{Evolutions of $\mathbf{L}_{\rm f}$ for $\delta = 0$ with $\l(\xi_e, \xi_\mu, \xi_\tau \r) = \l(-1.1, 1.6, 0.2 \r)$, $\theta = \l( \theta_{12}, \theta_{13}, \theta_{23} \r)$, and normal mass-hierarchy.
Colored/gray lines are with self-interaction turned-on/off.
}
\label{fig:self-on-off}
\end{center}
\end{figure}
For the numerical integrations of \eqs{eom-rho}{eom-rhobar}, we take the single mode approach as in Ref.~\cite{Barenboim:2016shh}.
The presence of self-interaction term in \eqs{eom-rho}{eom-rhobar} causes synchronized oscillations of density matrices, which becomes manifest for certain sets of $\xi_\alpha$.
However, as shown in Fig.~\ref{fig:self-on-off} for example, for the sets of parameters we are considering here, the evolution of density matrices is essentially the same as the case without self-interaction except the oscillatory feature which will be averaged out.
Hence, for simplicity and clarity of the result, we turn off the self-interaction.
Neutrino mass-square differences and mixing angles are taken as \cite{Agashe:2014kda,NOvA}
\bea
\Delta m_{21}^2 &\approx& 7.5 \times 10^{-5} \eV^2 
\\
|\Delta m_{31}^2| &\simeq& |\Delta m^2_{32}| =  2.67 \times 10^{-3} \eV^2
\eea
and 
\beq
\l( \theta_{12}, \theta_{13} \r) = \l( \frac{\pi}{5.5}, \frac{\pi}{20} \r), \quad \theta_{23} = \frac{\pi}{3.4}, \frac{\pi}{4}, \frac{\pi}{4.6}     
\eeq

\begin{figure*}[ht!]
\begin{center}
\includegraphics[width=0.47\textwidth]{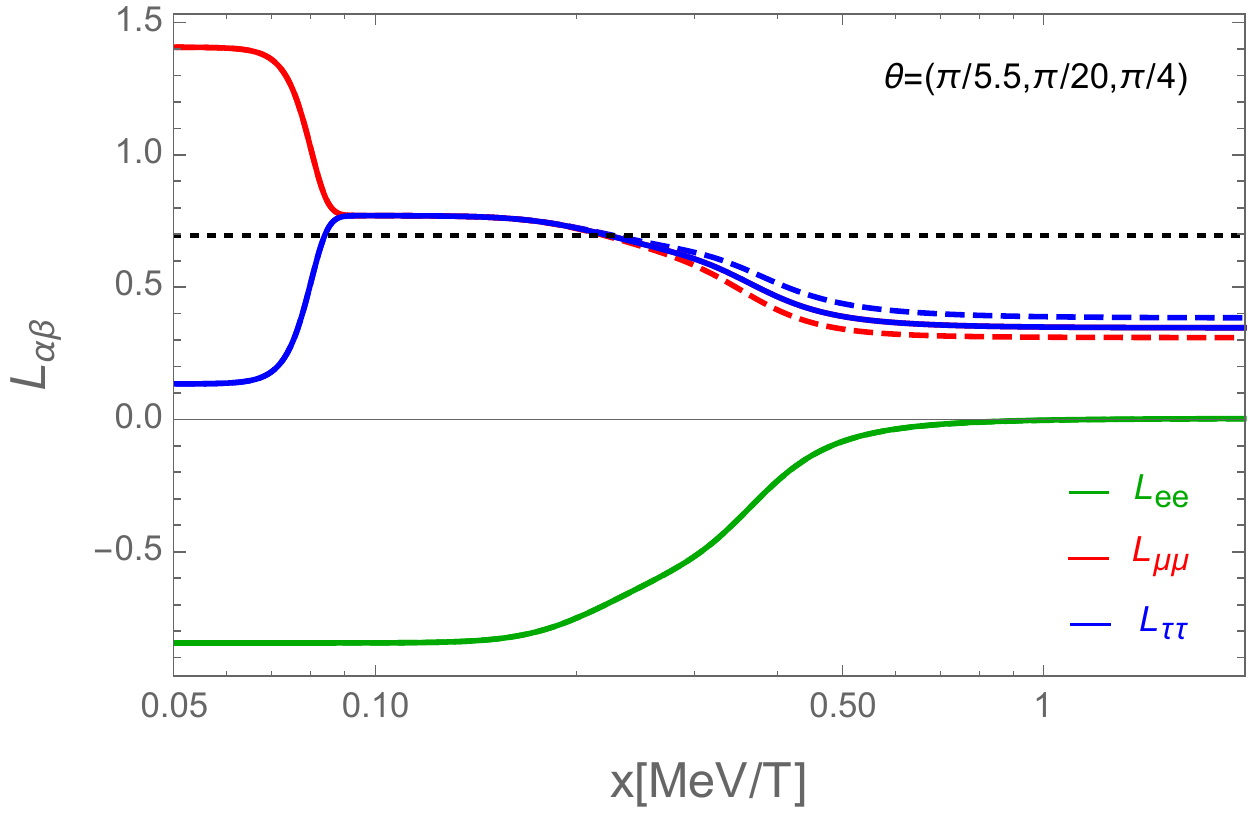}
\includegraphics[width=0.47\textwidth]{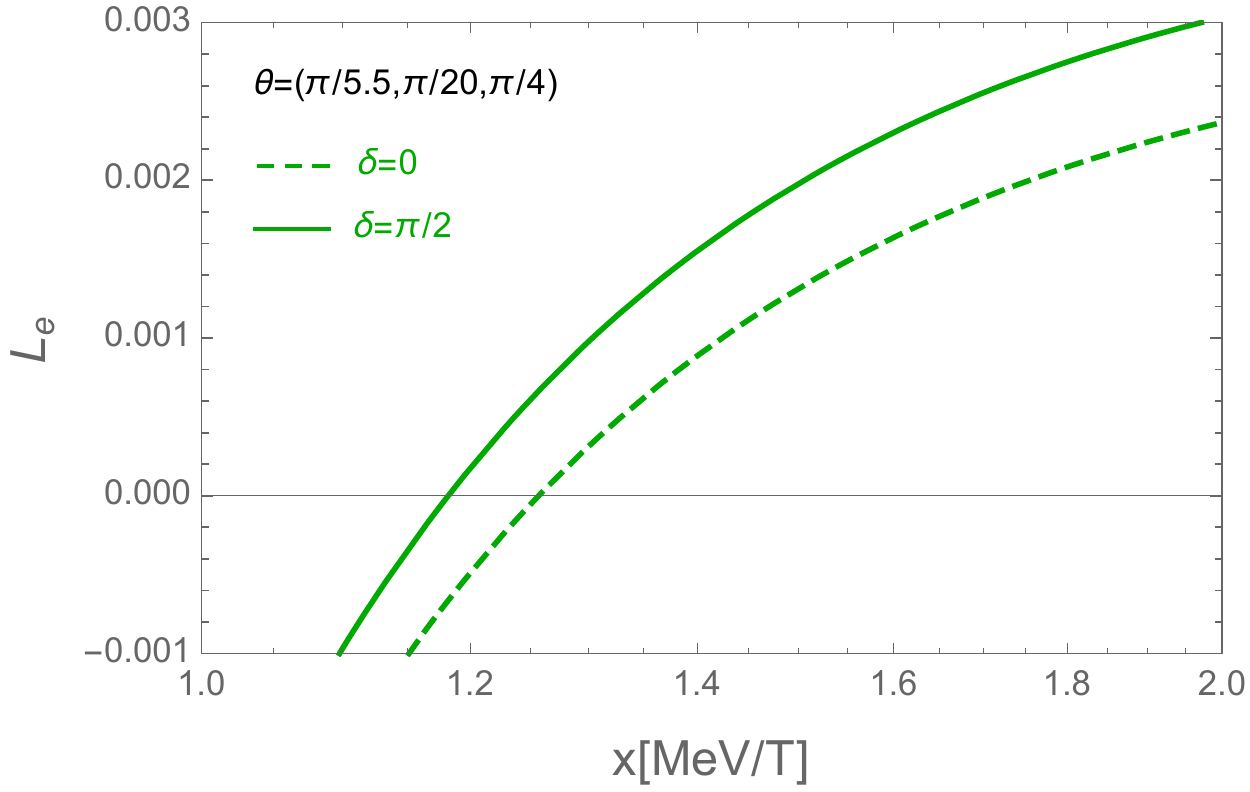}
\includegraphics[width=0.47\textwidth]{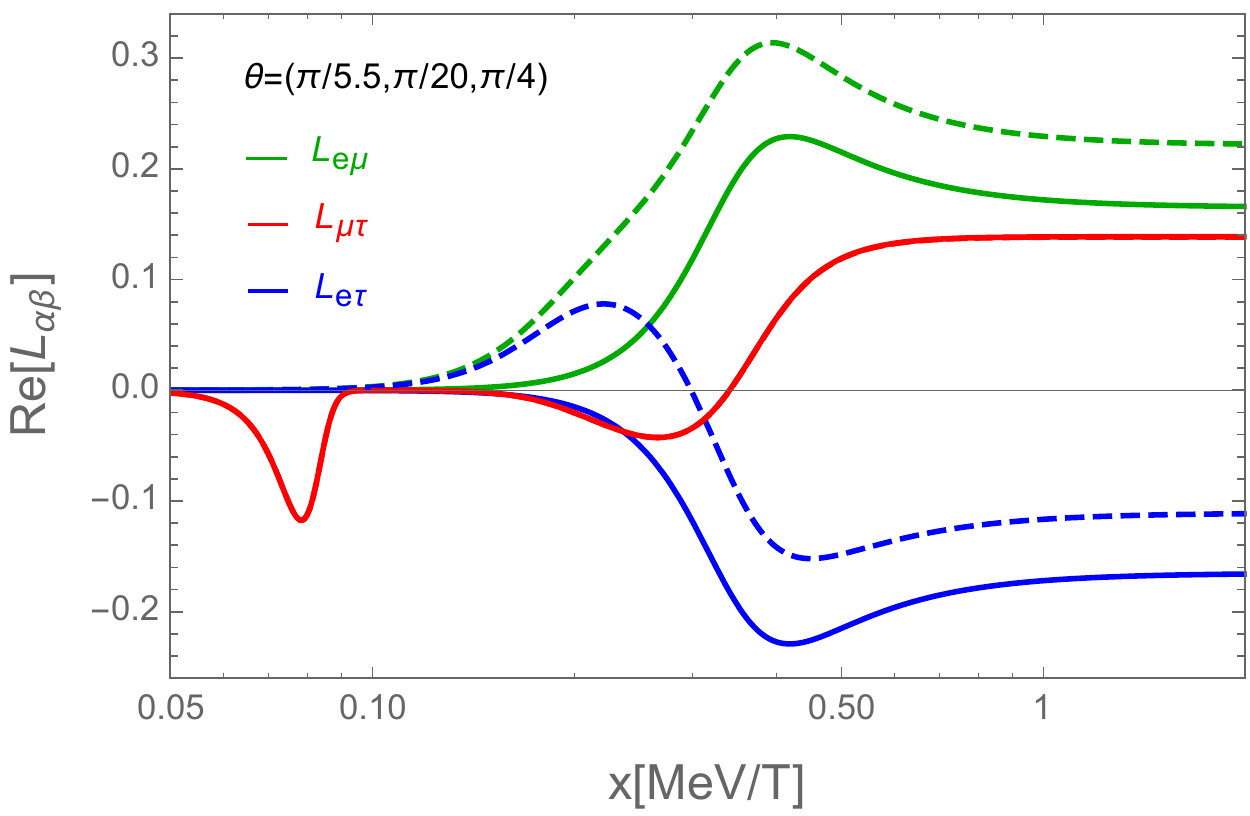}
\includegraphics[width=0.47\textwidth]{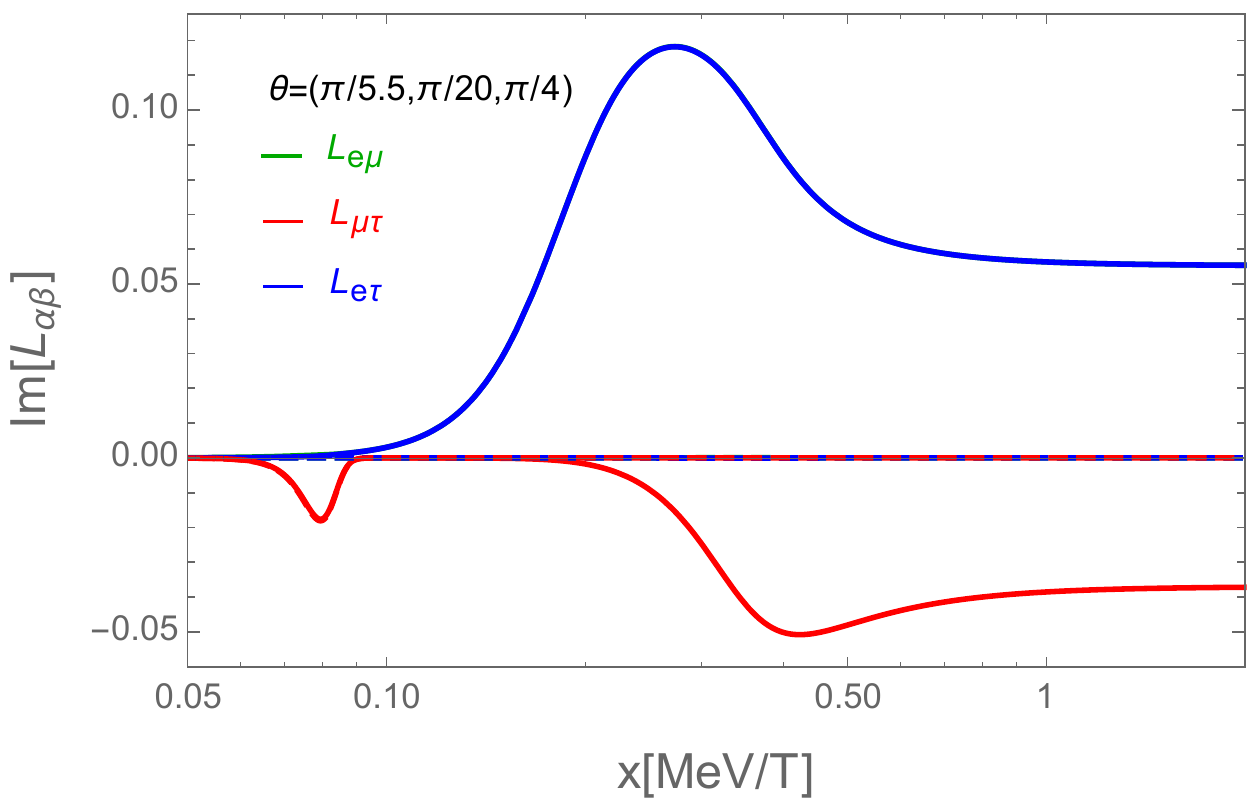}
\caption{Evolutions of $\mathbf{L}_{\rm f}$ for $\delta = 0$ (dashed lines) and $\pi/2$ (solid lines) with $\l(\xi_e, \xi_\mu, \xi_\tau \r) = \l(-1.1, 1.6, 0.2 \r)$, $\theta = \l( \theta_{12}, \theta_{13}, \theta_{23} \r)$, and normal mass-hierarchy.
\textit{Top-left}: Diagonal entries.
\textit{Top-right}: $L_e$.
\textit{Bottom-left}: Real components of off-diagonal entries. Dashed red line was overlapped by solid red line.
\textit{Bottom-right}: Imaginary components of off-diagonal entries. Green line was overlapped by blue line.
}
\label{fig:NH}
\end{center}
\end{figure*}
\begin{figure*}[ht!]
\begin{center}
\includegraphics[width=0.47\textwidth]{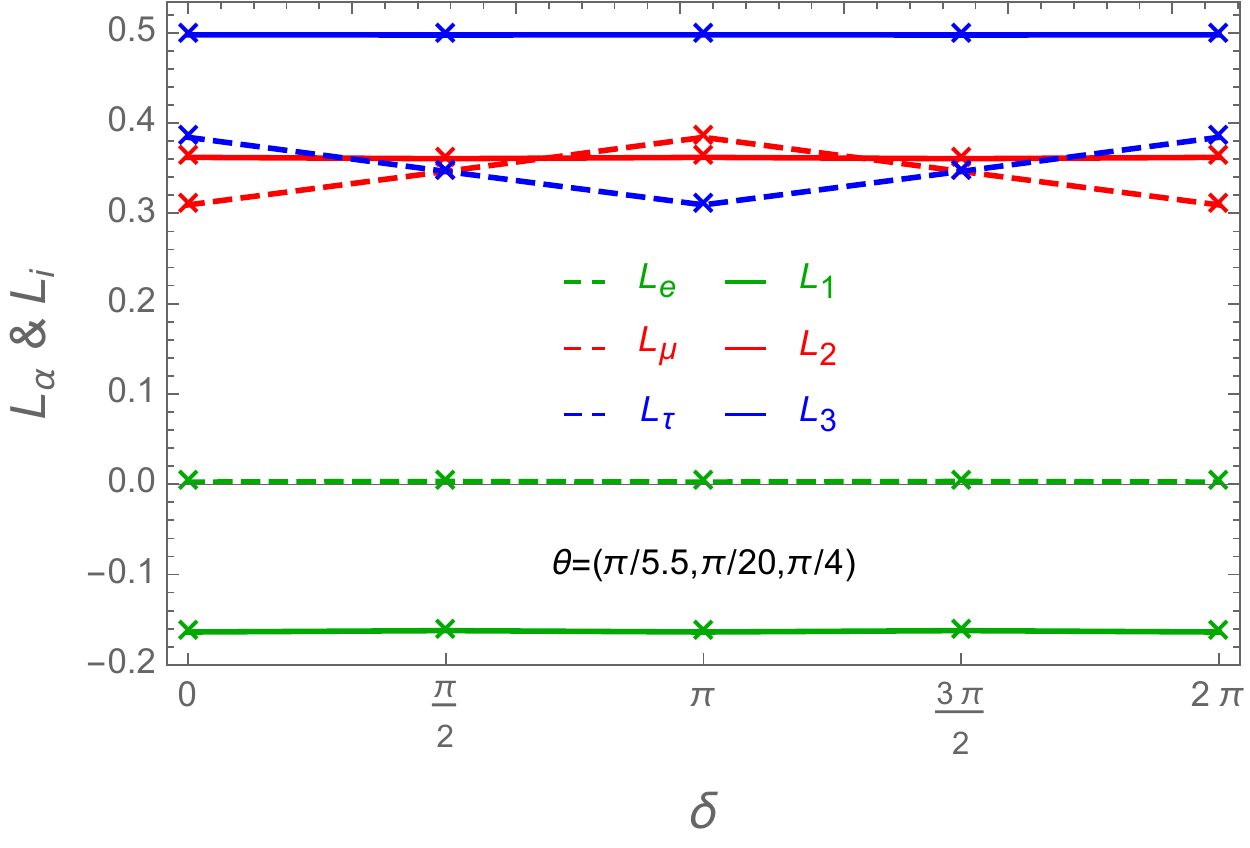}
\includegraphics[width=0.47\textwidth]{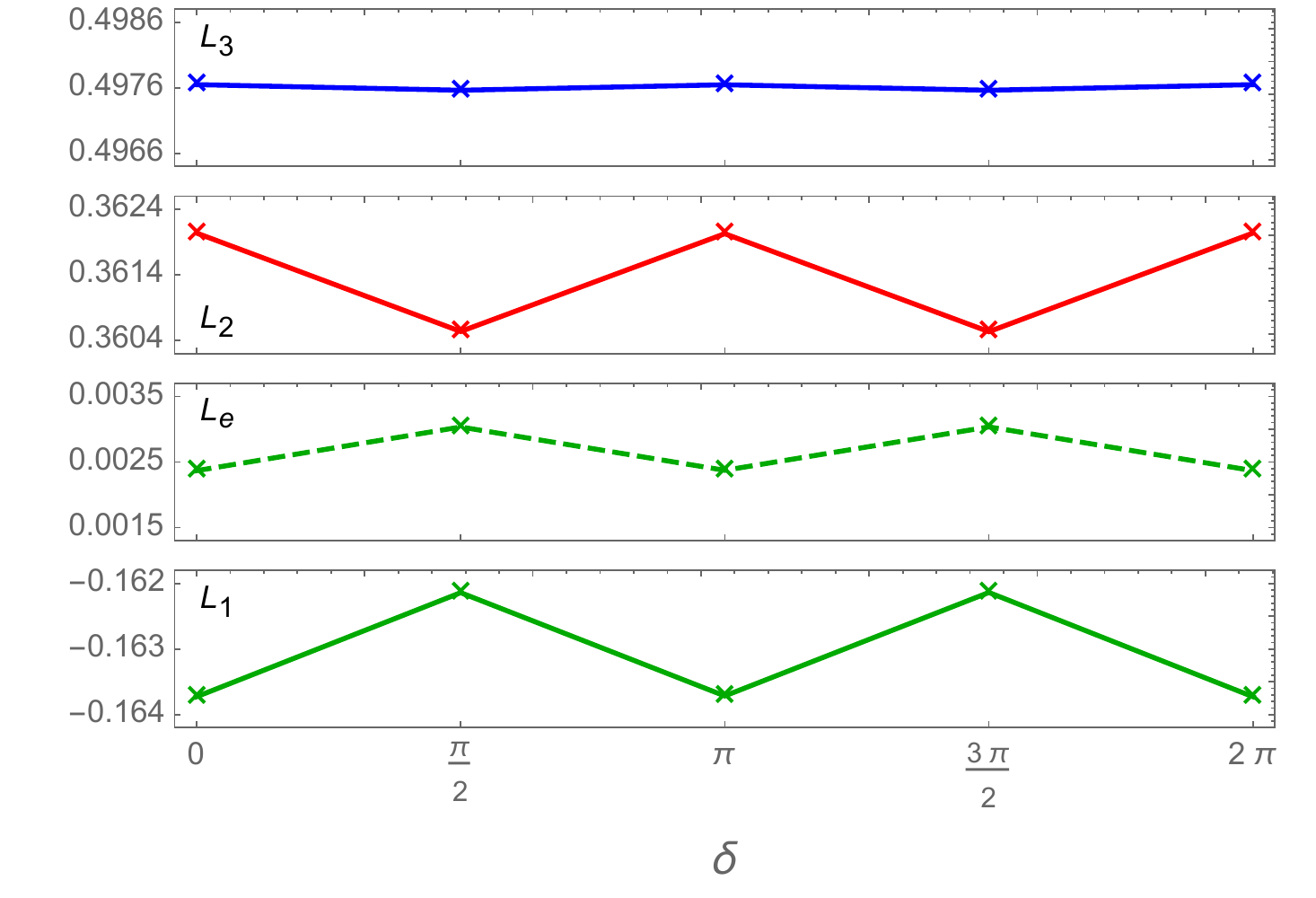}
\caption{The late time lepton number asymmetries of flavor- and mass-eigenstates (i.e., diagonal entries of $\mathbf{L}_{\rm f}$ and $\mathbf{L}_{\rm m}$ as a function of $\delta$ for normal mass-hierarchy.
The ``$\times$'' mark indicates actual data points obtained from numerical simulations.
\textit{Left/Right}: $L_\alpha$s and $L_i$s at different scales for clearer views of $\delta$-dependence.}
\label{fig:L-vs-delta}
\end{center}
\end{figure*}
In Fig.~\ref{fig:NH}, the evolution of the entries of $\mathbf{L}_{\rm f}$ are shown for $\delta = 0$ (dashed lines) and $\pi/2$ (solid lines).
$\xi_\alpha$ is chosen to have $L_e$ within the BBN constraint for $\delta=0$.
From the top-left panel where diagonal entries are shown, one can see that, if $\delta=0$, although $L_\mu$ and $L_\tau$ are completely equilibrated across $x \sim 0.06-0.07$ (or $T \sim 15 \MeV$) due to the large $\Delta m^2_{23}$ and $\theta_{23} \sim \pi/4$, they are separated across $x\sim 0.2$ (or $T\sim 5 \MeV$).
On the other hand, if $\delta=\pi/2$, such a separation does not take place, keeping $L_\mu=L_\tau$ once it is achieved.
This maintenance of equalization of $L_\mu$ and $L_\tau$ seems to be due to the specific patterns of \eqs{msq-eq}{CPV-equal}.        
For $\delta=\pi$, $L_\mu$ and $L_\tau$ exchange their positions relative to the case of $\delta=0$.
For $\delta=3\pi/2$, the result turns out to be the same as that of $\delta=\pi/2$, as could be  expected. 

In the same figure, the top-right panel shows how $L_e$ depends on $\delta$ in a limited (narrower) plot-range for a clear view of the difference.
Again, the results for $\delta=0$ and $\pi$ are the same, and so are $\delta=\pi/2$ and $3\pi/2$.
As $|\delta| \to \pi/2$, $L_e$ and $L_{\mu,\tau}$ become closer, maximizing the shift of $L_e$ at $\delta=\pi/2$ and $3\pi/2$ relative to the case of $\delta=0$ (or $\pi$).
For $\xi_\alpha \lesssim 1$, the shift is comparable to or smaller than the upper bound of $|L_e|$.
Such a change can be easily compensated by a different choice of initial $L_\alpha$.  
So,  practically it is difficult to constrain $\delta$ by the BBN constraint on $L_e$.
On the other hand, as the BBN constraint on $L_e$ becomes tighter, for a given $\delta$ the case of $L=0=L_e$ but $|\xi_{\mu,\tau}| \gtrsim 1$ as an initial configuration can be excluded since neutrino oscillations can cause non-zero $L_e$ at the final equilibrium, depending on $\delta$.

The bottom-left panel is showing the real off-diagonal entries of $\mathbf{L}_{\rm f}$.
While the changes of $L_{e\mu}$ and $L_{e\tau}$ are manifest, the change of $L_{\mu\tau}$ is negligible.
If $\delta=\pi$, relative to the case of $\delta=0$, the roles of $L_{e\mu}$ and $L_{e\tau}$ are reversed with opposite signs for both of them, but $L_{\mu\tau}$ is not affected.
For $\delta=3\pi/2$, the result is the same as the case of $\delta=\pi/2$.

The bottom-right panel is for the imaginary off-diagonal entries of $\mathbf{L}_{\rm f}$.   
In the panel, the line for $L_{e\mu}$ was completely overlapped with the line of $L_{e\tau}$.
Note that, while these off-diagonal entries become zero in case of $\delta=0 \ ({\rm or} \ \pi)$ (i.e., CP-conservation), they are non-zero for $\delta \neq 0 \ ({\rm or} \ \pi)$ (CP-violation). 
For $\delta = \pi$, the result is the same as the case of $\delta=0$.
For $\delta = 3\pi/2$, the result is the same as the case of $\delta=\pi/2$, but with signs flipped for both of $L_{\mu\tau}$ (for $x \gtrsim 0.1$) and $L_{e\mu, e\tau}$.

The matrix of the late time asymmetries shown in Fig.~\ref{fig:NH} turned out to be an Hermitian matrix containing imaginary off-diagonal entries.
We found that irrespective of CP-phase the diagonalization matrix is given by the PMNS matrix and the diagonalized matrix has only \text{real} entries, as it should be.
That is, again the asymmetries in mass-basis are obtained as \cite{Barenboim:2016lxv}
\beq
\mathbf{L}_{\rm m} = U_{\rm PMNS}^{-1} \mathbf{L}_{\rm f} U_{\rm PMNS}
\eeq
and $\mathbf{L}_{\rm m}$ is real and diagonal.
In Fig.~\ref{fig:L-vs-delta}, we show the late time asymmetries of neutrino flavor- and mass-eigenstates for the parameters of Fig.~\ref{fig:NH}.
As can be seen from the figure, even if CP-violation can cause sizeable changes in $L_\alpha$ (especially $L_\mu$ and $L_\tau$), its impact on $L_i$ causes changes of less than 1\% relative to the CP-conserving case. 

\begin{figure*}[ht!]
\begin{center}
\includegraphics[width=0.47\textwidth]{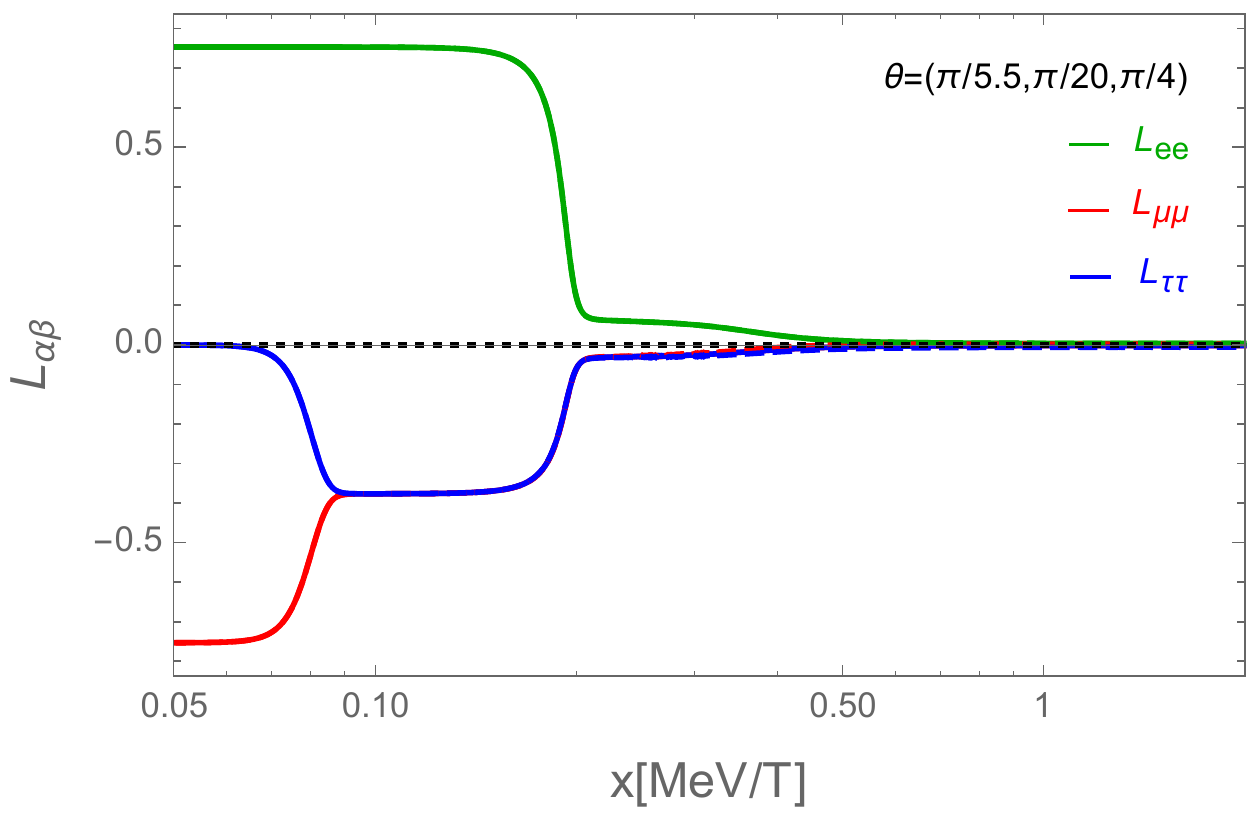}
\includegraphics[width=0.47\textwidth]{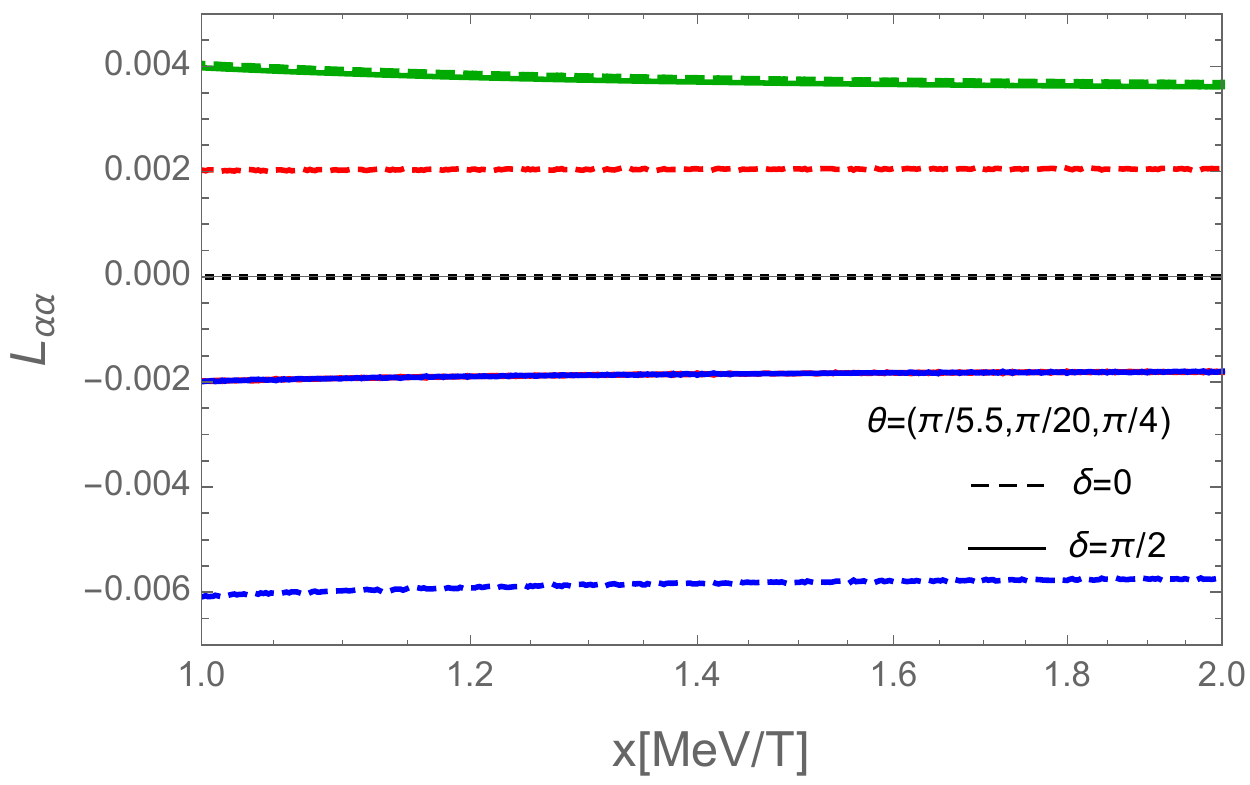}
\includegraphics[width=0.47\textwidth]{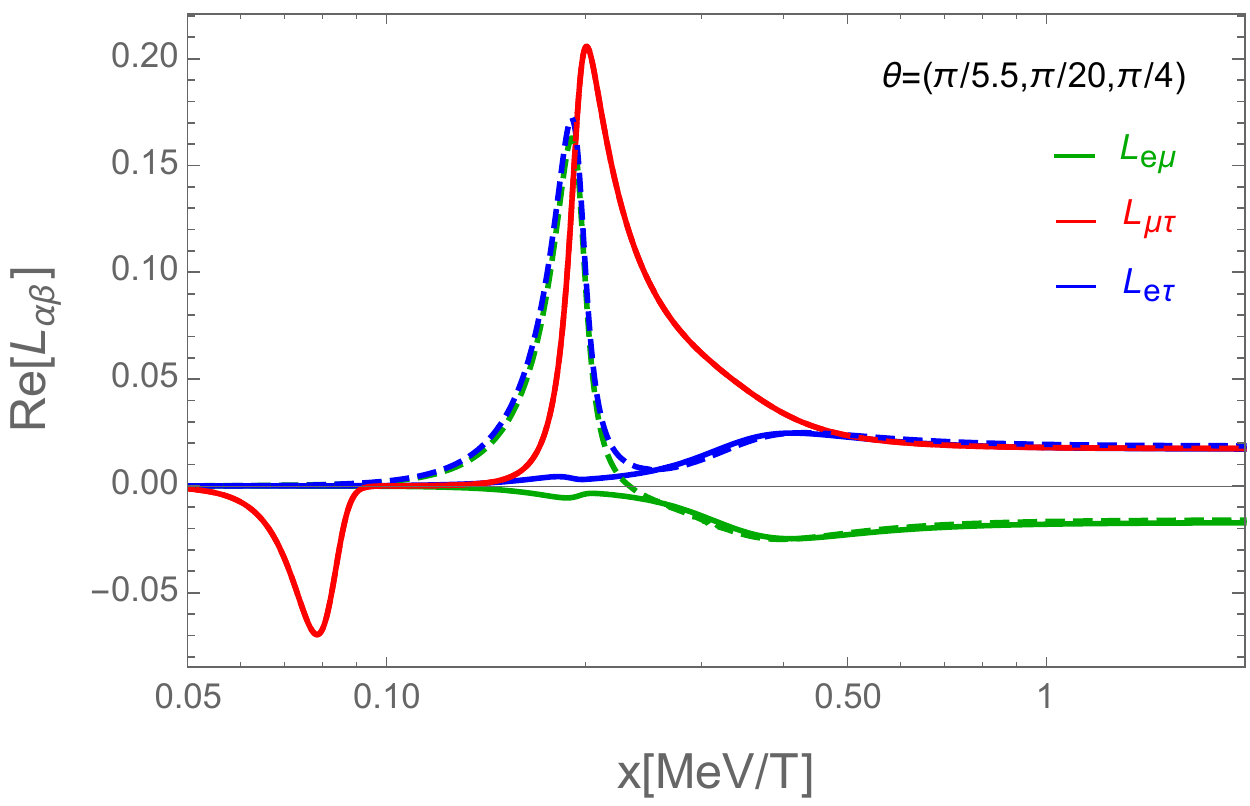}
\includegraphics[width=0.47\textwidth]{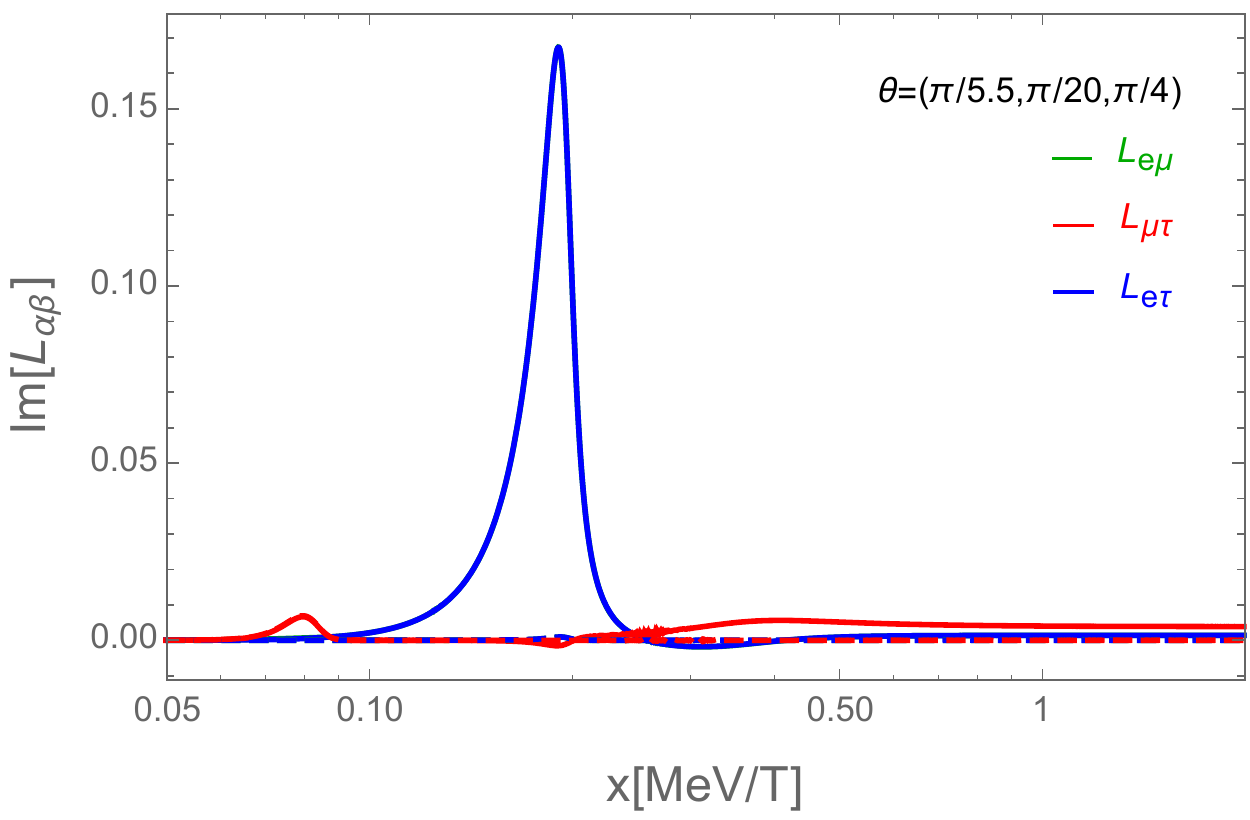}
\caption{Evolutions of $\mathbf{L}_{\rm f}$ for $\delta = 0$ (dashed lines) and $\pi/2$ (solid lines) with $\l(\xi_e, \xi_\mu, \xi_\tau \r) = \l(1, -1, 0 \r)$, $\theta = \l( \theta_{12}, \theta_{13}, \theta_{23} \r)$, and inverted mass-hierarchy.
\textit{Top-left}: Diagonal entries. For a specific $L_{\alpha\alpha}$,  the cases of both $\delta=0$ and $\pi/2$ are nearly overlapped with each other, and indistinguishable.
\textit{Top-right}: $L_{\alpha\alpha}$s at a scale different from top-left panel.
\textit{Bottom-left}: Real components of off-diagonal entries. Dashed red line was overlapped by solid red line.
\textit{Bottom-right}: Imaginary components of off-diagonal entries. Green line was overlapped by blue line.
}
\label{fig:IH}
\end{center}
\end{figure*}
\begin{figure*}[ht!]
\begin{center}
\includegraphics[width=0.47\textwidth]{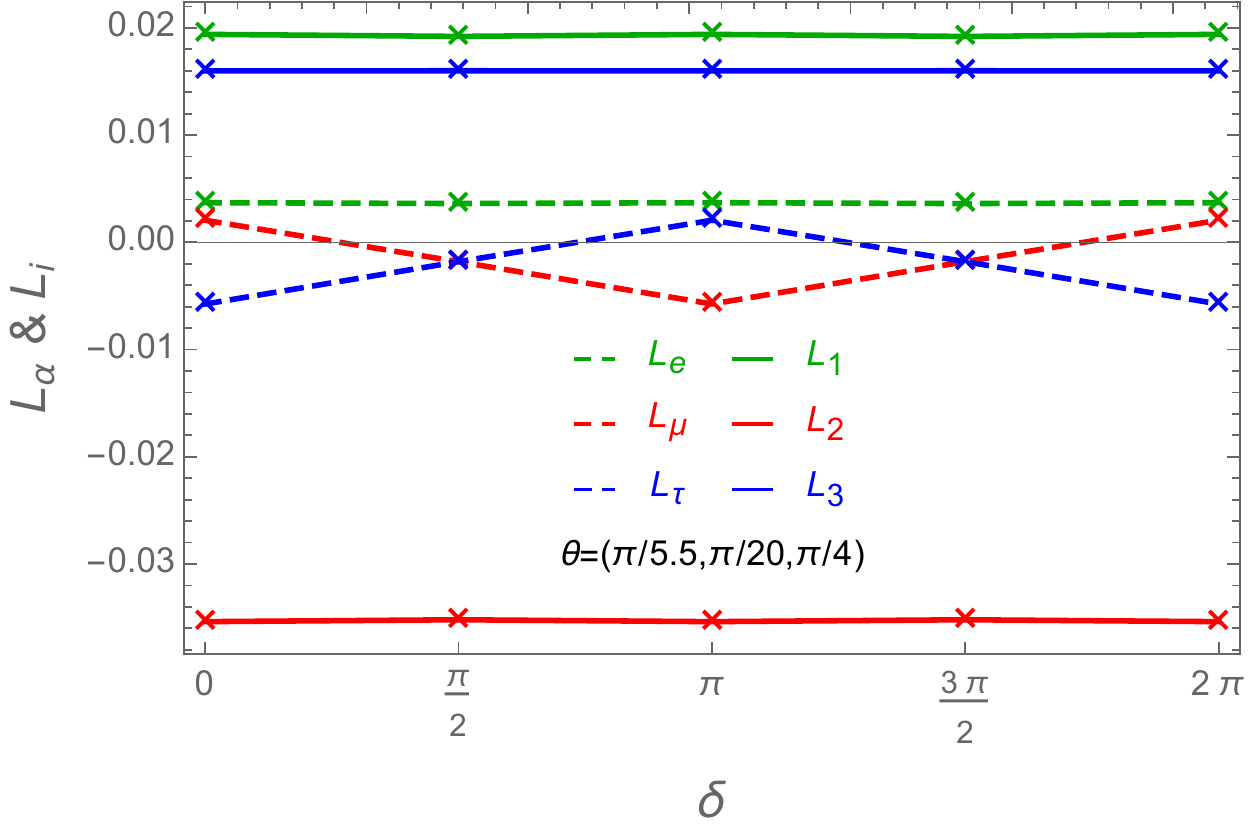}
\includegraphics[width=0.47\textwidth]{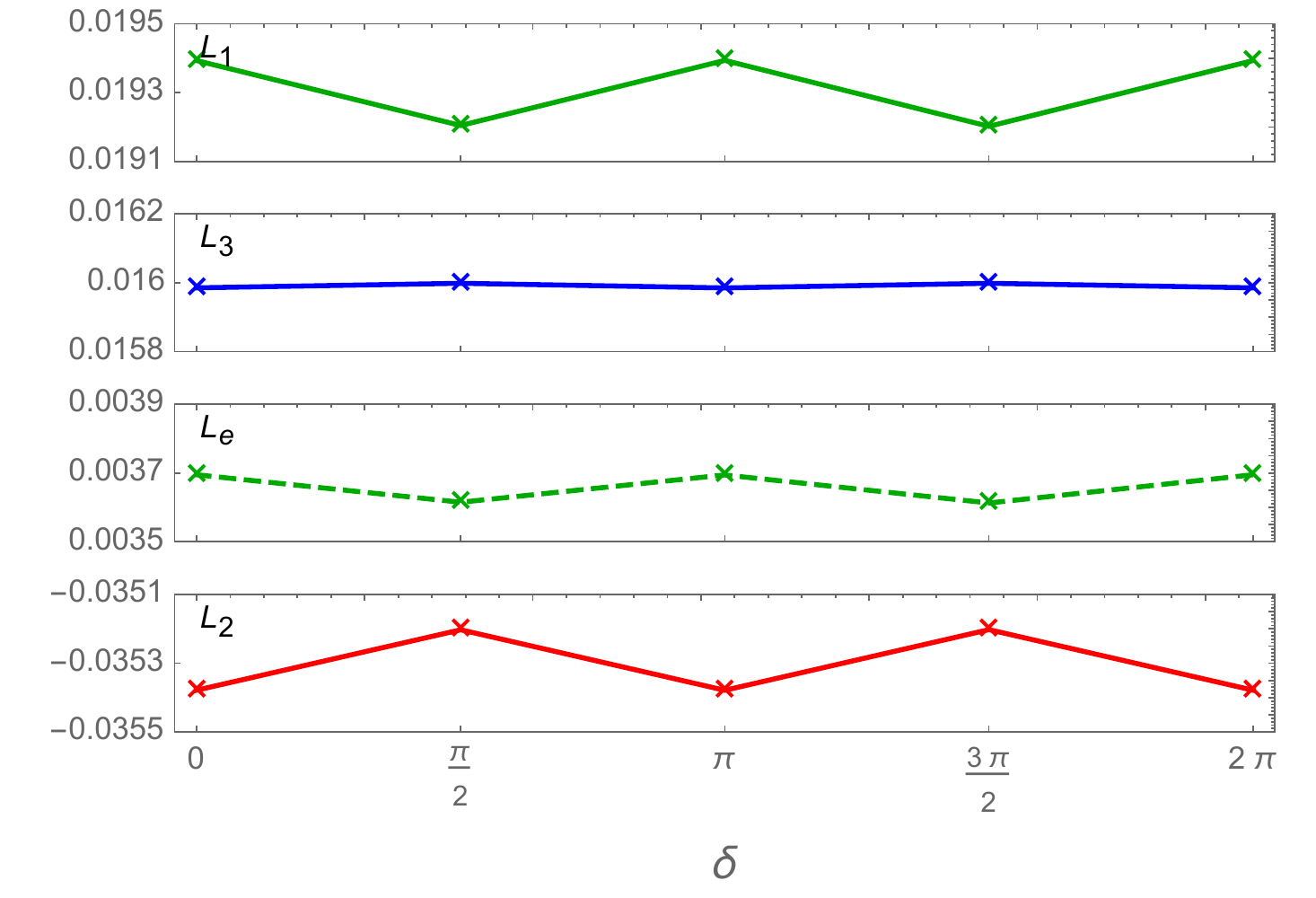}
\caption{The late time lepton number asymmetries of flavor- and mass-eigenstates (i.e., diagonal entries of $\mathbf{L}_{\rm f}$ and $\mathbf{L}_{\rm m}$ as a function of $\delta$ for inverted mass-hierarchy.
The ``$\times$'' mark indicates actual data points obtained from numerical simulations.
\textit{Left/Right}: $L_\alpha$s and $L_i$s at different scales for clearer views of $\delta$-dependence.}
\label{fig:IH-L-vs-delta}
\end{center}
\end{figure*}
The case of inverted mass-hierarchy is shown in Fig.~\ref{fig:IH} and~\ref{fig:IH-L-vs-delta}. 
As shown in the top-left panel of Fig.~\ref{fig:IH}, contrary to the case of normal mass hierarchy, in this case all $L_\alpha$s approach to the complete equalization across $x\sim0.5$ (or $T \sim 2 \MeV$). 
However, top-right panel shows that they do not reach complete equalization.
It may mean that, if BBN bound on $L_e$ becomes tighter in the future, the possibility of $\xi_\alpha \sim \mathcal{O}(1)$ as the initial degeneracy parameters of neutrino flavors at very high energy might be ruled out even if the total lepton number asymmetry is zero.

\begin{figure*}[ht!]
\begin{center}
\includegraphics[width=0.47\textwidth]{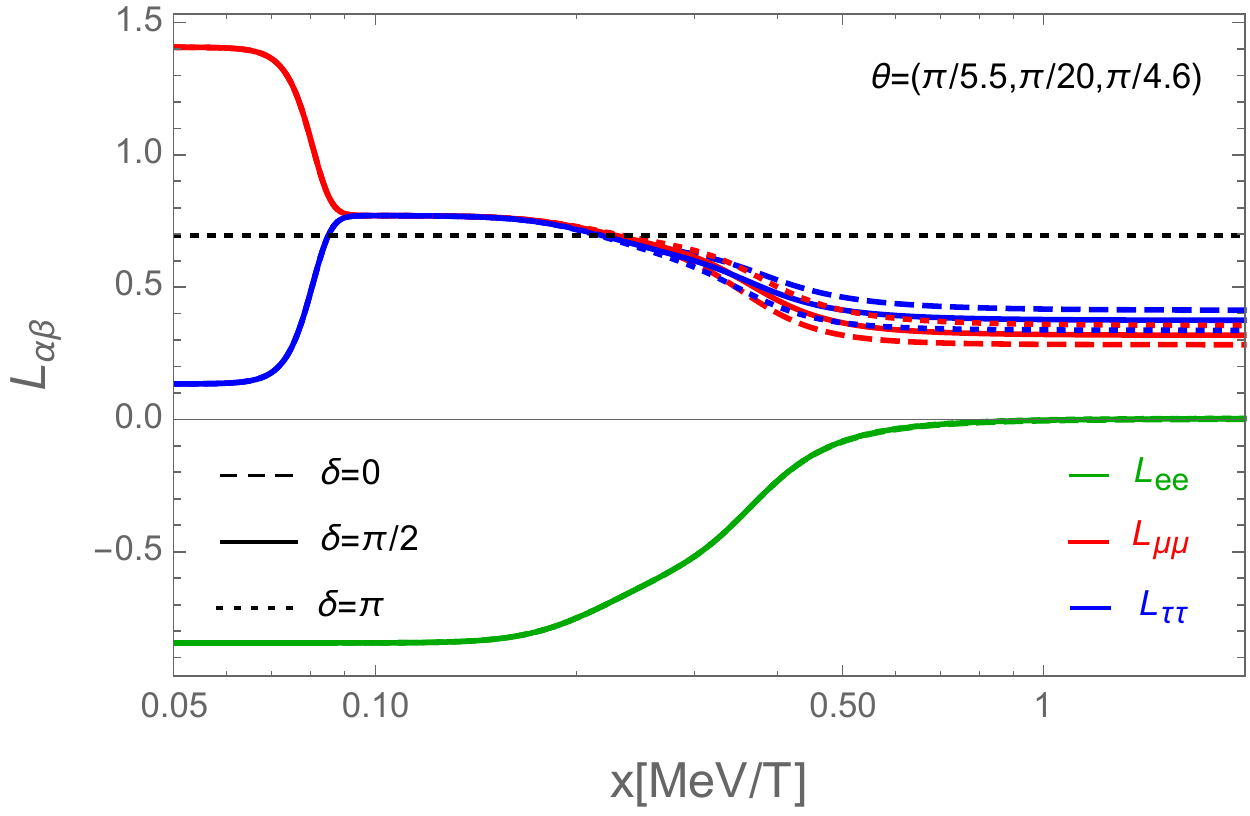}
\includegraphics[width=0.47\textwidth]{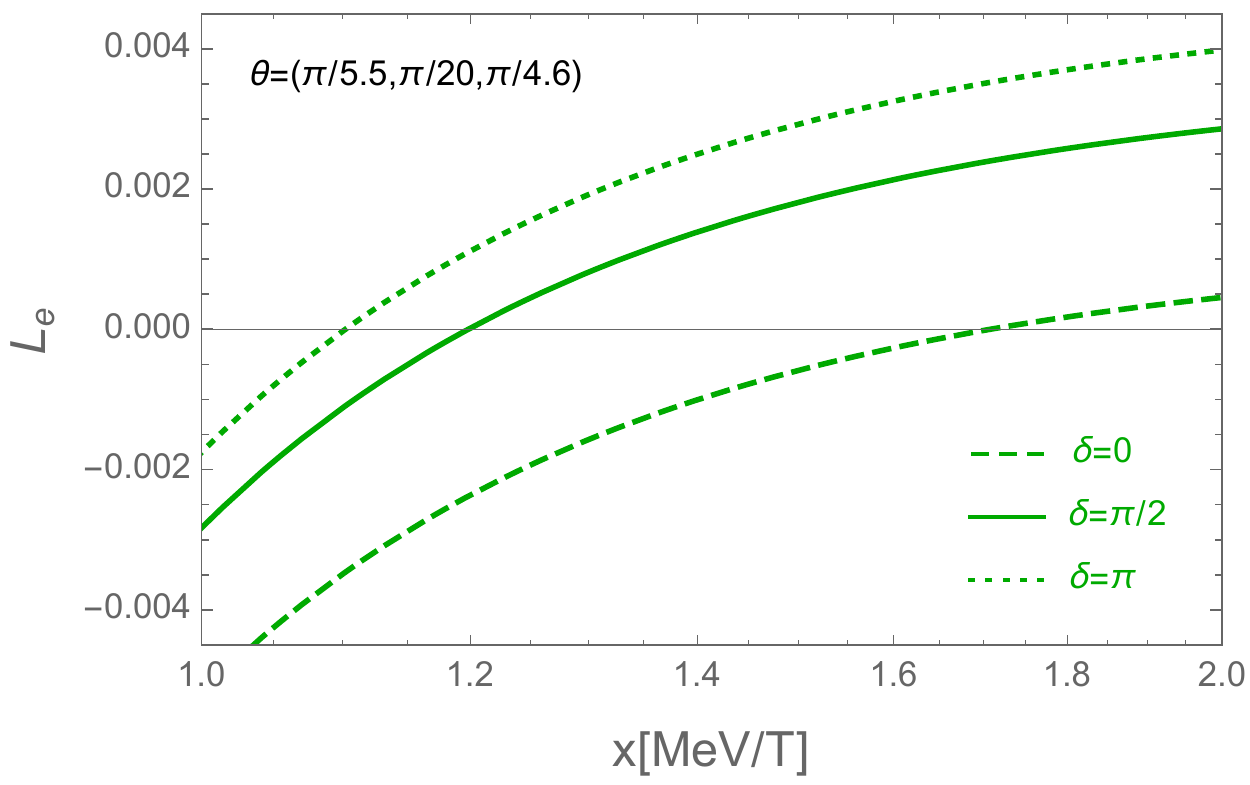}
\caption{Evolutions of $\mathbf{L}_{\rm f}$ for $\delta = 0$ (dashed lines), $\pi/2$ (solid lines), and $\pi$ (dotted lines) with $\l(\xi_e, \xi_\mu, \xi_\tau \r) = \l(-1.1, 1.6, 0.2 \r)$, $\theta = \l( \theta_{12}, \theta_{13}, \theta_{23} \r)$, and normal mass-hierarchy.
\textit{Left}: Diagonal entries.
\textit{Right}: $L_e$.
$\delta=3\pi/2$ gives the same result as $\delta=\pi/2$.
}
\label{fig:NH-theta23-pi4p6}
\end{center}
\end{figure*}
\begin{figure*}[ht!]
\begin{center}
\includegraphics[width=0.47\textwidth]{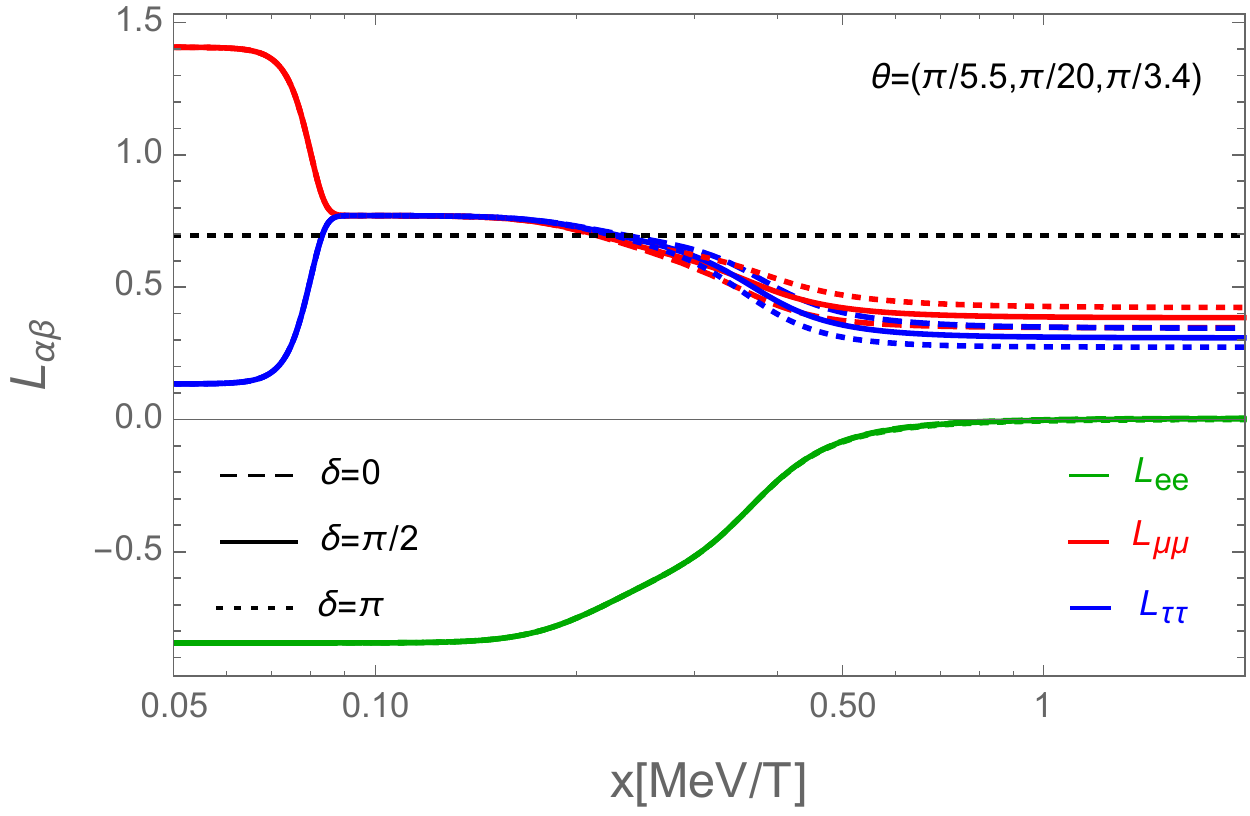}
\includegraphics[width=0.47\textwidth]{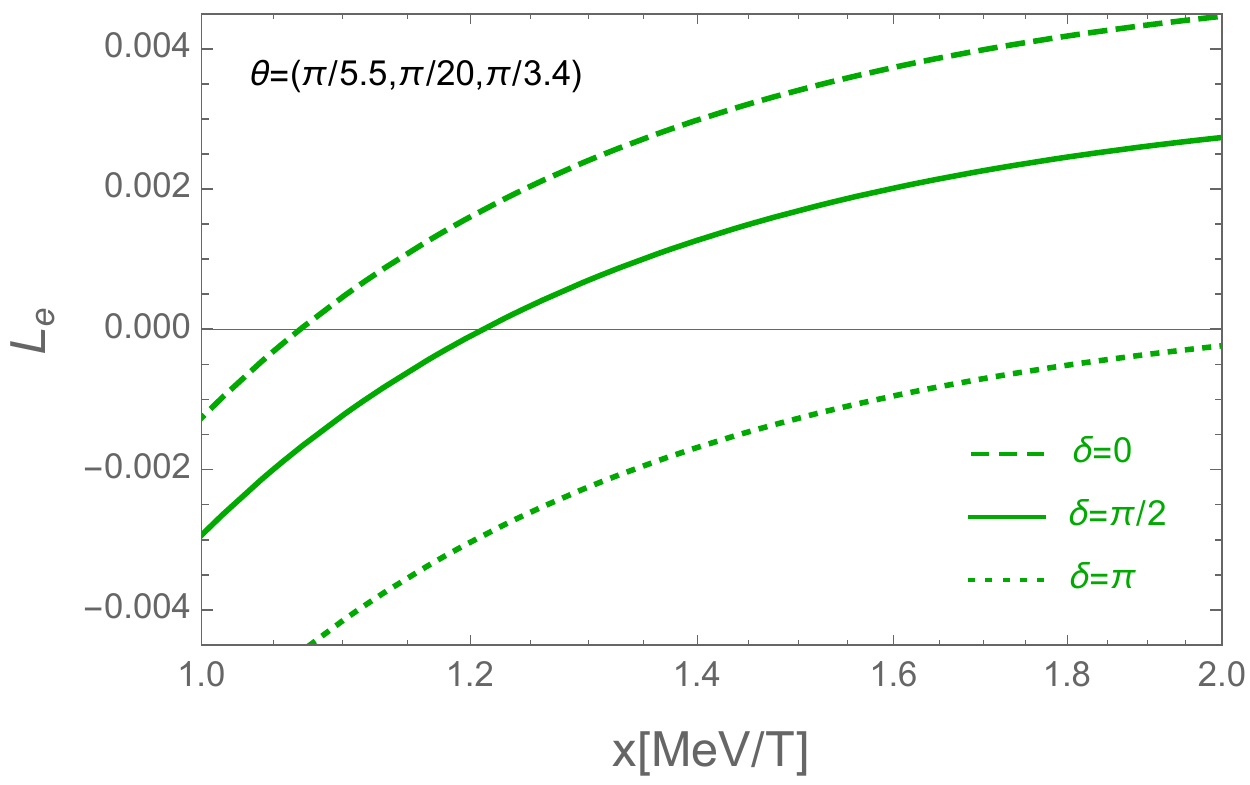}
\caption{The same as Fig.~\ref{fig:NH-theta23-pi4p6} but $\theta_{23}=\pi/3.4$ instead of $\pi/4.6$.
Again, $\delta=3\pi/2$ gives the same result as $\delta=\pi/2$.
}
\label{fig:NH-theta23-pi3p4}
\end{center}
\end{figure*}
Comparing top-right panel to bottom panels, we found that the real components of off-diagonal entries are larger than diagonal entries by a factor of a few. 
This fact results in mass-basis asymmetries much larger than ones in flavor basis, as shown in Fig.~\ref{fig:IH-L-vs-delta} where $\delta$-dependence of asymmetries in both bases is depicted.
Similarly to the case of normal mass-hierarchy, CP-phase can cause sizable changes in $L_\mu$ and $L_\tau$, but the change of $L_e$ is at most of $\mathcal{O}(1)$\%.
The smallness of the change seems to be from the smallness of the late time $L_{\mu,\tau}$.
The changes in the asymmetries in mass-basis is less than 1\% as the case of normal hierarchy.
Also, the pattern of the evolutions of $\mathbf{L}_{\rm f}$ for $\delta=\pi$ ($3\pi/2$) relative to the case of $\delta=0$ ($\pi/2$) is the same as the case of normal mass-hierarchy.

For a normal mass hierarchy, the late time asymmetries at the final equilibrium depend on the mixing angles critically.
For example, in the case of $\delta=0$, for a given set of initial flavor asymmetries (of $L_e \neq L_{\mu,\tau}$) which end up in a value of $L_e$ satisfying the BBN bound for a set of $\theta_{ij}$, the smaller $\theta_{12}$ or $\theta_{13}$ is, the larger $|L_\mu-L_e|$ or $|L_{\tau}-L_e|$ becomes, respectively, as expected.
On the other hand, the smaller $\theta_{23}$ is, the larger $|L_\mu-L_\tau|$ becomes.
The $\delta$-dependence for different values of $\theta_{23}$ is shown in Figs.~\ref{fig:NH-theta23-pi4p6} and~\ref{fig:NH-theta23-pi3p4}.
In Fig.~\ref{fig:NH-theta23-pi4p6}, we find that at the final equilibrium,  if $L_\tau > L_\mu \gg |L_e|$ for $\delta=0$ with a given set of $\theta_{ij}$, non-zero $\delta$ makes $L_\tau$ and $L_\mu$ shifted toward each other by an equal amount in addition to an overall shifting toward $L_e$, resulting in a shift of $L_e$ to keep the conservation of the total asymmetry.
Such a shift of $L_e$ is maximized when $L_\mu=L_\tau$.
Hence, the largest effect of non-zero $\delta$ appears when $L_\mu$ becomes the closest to $L_\tau$ as happens when $\delta=\pi$. 
If $L_\mu = L_\tau$ for a certain set of $\theta_{ij}$ with $\delta=0$, non-zero $\delta$ pushes $L_e$ away from $L_{\mu,\tau}$, as shown in Fig.~\ref{fig:NH-theta23-pi3p4}.
The changes in the final lepton number asymmetries of flavor- and mass-eigenstates relative to the case of $\theta_{23}=\pi/4$ are minor and do not affect our conclusions.

\section{Conclusions}

In this paper, we revisited the effect of CP-violation (i.e., nonzero Dirac CP-phase ($\delta$) in the PMNS matrix) on the neutrino lepton number asymmetries in both flavor- and mass-basis by solving the evolution equations of neutrino/anti-neutrinos density matrices in a simplified single mode approach as taken on our recent work, Ref.~\cite{Barenboim:2016shh}.

In the evolution equations of neutrino density matrices, the effect of CP-violation appears in the mass-square matrix $\mathbf{M}^2_{\rm f}$ in the flavor basis.
In particular, the effect is dominated by $\delta$-dependence of the off-diagonal entries of $\mathbf{M}^2_{\rm f}$ associated with $\nu_e-\nu_\mu$ and $\nu_e-\nu_\tau$ mixings.
For $\theta_{23} \sim \pi/4$ as one of the mixing angles of the PMNS matrix, it turned out that the $\delta$-dependence of $L_\mu$ and $L_\tau$ at the final equilibrium is nearly equal and opposite.
As a result, the change of $L_e$ appears to be much smaller than the change of $L_{\mu,\tau}$ since the total lepton number asymmetry should be conserved.

The effect of CP-violation turned out to be maximized for a $\delta$ such hat the values of $L_\mu$ and $L_\tau$ are the closest or farthest from each other. 
In normal mass hierarchy, the change of $|L_{\mu,\tau}|$ is about a few tens of \% at most.
However, the change of $|L_e|$ is within the range of the upper-bound set by BBN as long as $\xi_\alpha \lesssim 1$ with $\xi_\alpha$ being the initial degeneracy parameters of neutrino flavors.
In inverted hierarchy, one finds $|L_e| \sim |L_\mu| \sim |L_\tau|$ at late time.
Hence, $L_{\mu,\tau}$ is constrained to be small to satisfy the BBN constraint on $L_e$.
CP-violation in this case can significantly increase  $L_{\mu,\tau}$, with a raise up to more than 100\%, but the change in $L_e$ appears to be negligible.

Contrary to the cases of flavor eigenstates, the effect of CP-violation on asymmetries of mass-eigenstates, the ones relevant to CMB,  turns out to be practically negligible.
It causes sub-\% changes of asymmetries which are difficult to be explored in current or future experiments, even for initial asymmetries as large as $\xi_\alpha \sim 1$.

\section{Acknowledgements}
The authors are grateful to William H. Kinney for helpful conversations. 
They also acknowledge support from the MEC and FEDER (EC) Grants SEV-2014-0398 and FPA2014-54459 and the Generalitat Valenciana under grant PROMETEOII/2013/017. This project has received funding from the European Union's Horizon 2020
research and innovation programme under the Marie Sklodowska-Curie grant
Elusives ITN agreement No 674896  and InvisiblesPlus RISE, agreement No 690575. 



\begin{thebibliography}{99}

\bibitem{Pontecorvo:1957cp} 
  B.~Pontecorvo,
  Sov.\ Phys.\ JETP {\bf 6}, 429 (1957)
  [Zh.\ Eksp.\ Teor.\ Fiz.\  {\bf 33}, 549 (1957)].

\bibitem{Maki:1962mu} 
  Z.~Maki, M.~Nakagawa and S.~Sakata,
  Prog.\ Theor.\ Phys.\  {\bf 28}, 870 (1962).
  doi:10.1143/PTP.28.870



\bibitem{Bilenky:1980cx} 
  S.~M.~Bilenky, J.~Hosek and S.~T.~Petcov,
  Phys.\ Lett.\ B {\bf 94}, 495 (1980).

\bibitem{Cabibbo:1977nk} 
  N.~Cabibbo,
  Phys.\ Lett.\ B {\bf 72}, 333 (1978).

\bibitem{Barger:1980jm} 
  V.~D.~Barger, K.~Whisnant and R.~J.~N.~Phillips,
  Phys.\ Rev.\ Lett.\  {\bf 45}, 2084 (1980).

\bibitem{Bernabeu:2010rz} 
  J.~Bernabeu {\it et al.},
  arXiv:1005.3146 [hep-ph].

\bibitem{Gava:2010kz} 
  J.~Gava and C.~Volpe,
  Nucl.\ Phys.\ B {\bf 837}, 50 (2010)




\bibitem{Barenboim:2016shh} 
  G.~Barenboim, W.~H.~Kinney and W.~I.~Park,
  arXiv:1609.01584 [hep-ph].

\bibitem{Barenboim:2016lxv} 
  G.~Barenboim, W.~H.~Kinney and W.~I.~Park,
  arXiv:1609.03200 [astro-ph.CO].




\bibitem{Sigl:1992fn} 
  G.~Sigl and G.~Raffelt,
  Nucl.\ Phys.\ B {\bf 406}, 423 (1993).
  doi:10.1016/0550-3213(93)90175-O


\bibitem{Pantaleone:1992eq} 
  J.~T.~Pantaleone,
  Phys.\ Lett.\ B {\bf 287}, 128 (1992).
  doi:10.1016/0370-2693(92)91887-F


\bibitem{Dolgov:2002ab} 
  A.~D.~Dolgov, S.~H.~Hansen, S.~Pastor, S.~T.~Petcov, G.~G.~Raffelt and D.~V.~Semikoz,
  Nucl.\ Phys.\ B {\bf 632}, 363 (2002)
  doi:10.1016/S0550-3213(02)00274-2
  [hep-ph/0201287].


\bibitem{Abazajian:2002qx} 
  K.~N.~Abazajian, J.~F.~Beacom and N.~F.~Bell,
  Phys.\ Rev.\ D {\bf 66}, 013008 (2002)
  doi:10.1103/PhysRevD.66.013008
  [astro-ph/0203442].




\bibitem{Agashe:2014kda} 
  K.~A.~Olive {\it et al.} [Particle Data Group Collaboration],
  Chin.\ Phys.\ C {\bf 38}, 090001 (2014).
  doi:10.1088/1674-1137/38/9/090001

\bibitem{NOvA}
https://www-nova.fnal.gov/







%
%
%
%
%
%
%
%
%
%
%
%
%
%
%
%
%
%
%
%
%
%
%
%
%
%
%
%
%
%
%
%
%
%
%
%
%
%
%
%
%
%
%
%
%
%
%
%
%
%
%
%
%
%
%
%
%
%
%
%
%
%
%
%
%


\end{thebibliography}
\end{document}